\documentclass[preprintnumbers,aps,nofootinbib,amsmath,amssymb,12pt,floatfix,epsfig,showpacs,showkeys]{revtex4}
\usepackage{lineno,hyperref}
\usepackage{epsfig}
\usepackage{epstopdf}
\usepackage{graphicx}
\usepackage[dvips]{color}

\begin{document}

\title{A new neutrino source for the study of the solar neutrino physics
in the vacuum-matter transition region}

\author{Jae Won Shin}
\affiliation{Department of Physics,
Soongsil University, Seoul 156-743, Korea}
\author{Myung-Ki Cheoun}
\email{cheoun@ssu.ac.kr}
\affiliation{Department of Physics,
Soongsil University, Seoul 156-743, Korea}

\author{Toshitaka Kajino}
\affiliation{Division of Theoretical Astronomy, 
National Astronomical Observatory 
of Japan, Mitaka, Tokyo 181-8588, Japan \\
and Department of Astronomy, Graduate School of Science, University of 
Tokyo,  Hongo, Bunkyo-ku, Tokyo 113-0033, Japan}

\date{Jan. 12, 2016}

\begin{abstract}
Production of a neutrino source through proton induced reaction
is studied by using the particle transport code, GEANT4.
Unstable isotope such as $^{27}$Si can be produced
when  $^{27}$Al target is bombarded by
15 MeV energetic proton beams.
Through the beta decay process of
the unstable isotope,
a new electron-neutrino source in the 1.0 $\sim$ 5.0 MeV energy range
is obtained.
Proton induced reactions are simulated with
JENDL High Energy File 2007 (JENDL/HE-2007) data
and other nuclear data.
For radioactive decay processes,
we use ``G4RadioactiveDecay" model based on
the Evaluated Nuclear Structure Data File (ENSDF).
We suggest target systems required
for future's solar neutrino experiments,
in particular, for the vacuum-matter transition region.
As for the detection system of the new neutrino source,
we evaluate reaction rates for available radiochemical detectors
and LENA type scintillator detector.
Possibility of detecting sterile neutrinos is also discussed.
\end{abstract}


\maketitle

\section{Introduction}

The neutrino oscillation mechanism which was firstly suggested
by Pontecorvo \cite{nu_oscill_1, nu_oscill_2, nu_oscill_3, nu_oscill_4, nu_oscill_5}
plays important roles to explain the discrepancies among
the measured solar neutrino flux and/or the theoretical one,
so-called solar neutrino problem.
Since the first observation of the effects of neutrino oscillation
was done in Homestake Experiments \cite{oscil_exp_1},
many experiments,
such as KamioKanDe \cite{Kamio_1} and SNO \cite{SNO_1} facilities,
had confirmed the neutrino oscillation through more refined
and advanced detection.

In addition to the oscillation in vacuum,
the recent experimental data at Borexino \cite{Borexino_1},
which measured $pep$ neutrino in 1.0 $\sim$ 1.5 MeV,
point out to need to consider Mikheyev-Smirnov-Wolfenstein (MSW)
\cite{msw_1, msw_2, msw_3, msw_4}
large mixing angle (LMA) solutions.
The MSW model predicts a transition from vacuum-dominated to
matter-enhanced oscillations relevant to the metallicity problem in the solar model, but experimental understanding
is still insufficient in this energy region.

One of the future open issues
related to the vacuum-matter transition region
in the solar neutrino physics is
the determination of the electron-neutrino survival probability
P(${\nu}_{e}$ ${\to}$ ${\nu}_{e}$) in that region \cite{Review_nu_1},
which is also closely related to questions
such as
existence of sterile neutrinos \cite{q_sterile1, q_sterile2}
and/or the non-standard neutrino interactions (NSI) \cite{q_NSI},
roles of CNO cycle in the Sun (metallicity problem) \cite{q_metal_1, q_metal_2},
etc.

For this reason,
it becomes of importance to detect in detail the solar neutrino in the transition region.
Moreover, it would be more efficient if we have the controllable neutrino source in the energy region,
which may enable us to extract both the total number and the energy distribution
of neutrinos in a more accurate way, with the more elaborate detection system.
In this work, we propose an accelerator based new artificial electron-neutrino source
for the experiments of the vacuum-matter transition region.
By adjusting the incident proton energy,
we can produce a specific unstable isotope
as an efficient electron-neutrino source.
Unstable isotope, $^{27}$Si, is our main neutrino source,
which can be produced through $^{27}$Al(p,n)$^{27}$Si reaction
and emits electron-neutrinos through radioactive decay processes.
In this case,
the neutrinos can have the energy similar to the transition region.

We outline the paper in the following way.
In Sec. \ref{sec2},
we summarize the simulation method used in this work.
Benchmarking simulations for $^{27}$Al(p,n)$^{27}$Si reaction,
calculations for $^{27}$Si yields and
energy spectra of electron-neutrinos from decay of $^{27}$Si
are presented in Sec. \ref{sec3}.
Electron-neutrino detections, event rate of the scattered electron and
possibility of detecting the sterile neutrino are discussed in Sec. \ref{sec4} and \ref{sec5}, respectively.
The summary is given in Sec. \ref{sec6}.

\section{Simulation method
\label{sec2}}

As an electron-neutrino source,
we consider $^{27}$Si isotope in this work,
which can be produced through $^{27}$Al(p,n)$^{27}$Si reaction
with a threshold energy E$_{th}$ of 5.803 MeV.
To evaluate $^{27}$Si yields produced by proton beams
on $^{27}$Al target,
we use the particle transport code,
GEANT4 (GEometry ANd Tracking) v10.1 \cite{g4n1, g4n2},
which is a tool kit
that allows for microscopic Monte Carlo simulations of
particles interacting with materials.

For proton inelastic scattering,
four different hadronic models such as
``G4BertiniCascade" \cite{BERTIref3},
``G4BinaryCascade" \cite{BICref1},
``G4Precompound" \cite{PRECOMref1} and
``G4INCLCascade" \cite{inclPref}
are available in GEANT4 (v10.1).
To check the validity of the models,
we first perform simulations of $^{27}$Si production
by p + $^{27}$Al reaction and
compare the calculated results with the experimental data
taken from the EXFOR database \cite{exfor}.
For brevity,
we refer to GEANT4 simulations with ``G4BertiniCascade",
``G4BinaryCascade", ``G4Precompound" and ``G4INCLCascade" as
``G4BERTI", ``G4BC", ``G4PRECOM" and ``G4INCL", respectively.
Detailed descriptions and additional information
of the above mentioned models
are described well in the Physics Reference Manual \cite{g4_physRef}
and Refs. \cite{G4HadPhy1, G4HadPhy2}.

Recently,
discrepancies between the experimental data and
the simulation results obtained from hadronic models of GEANT4
in the low energy region were discussed in detail in Refs. \cite{g4ce, g4dpn}.
For more accurate simulations,
data-based hadronic models for GEANT4 were developed
by incorporating the Evaluated Nuclear Data File (ENDF/B-VII.1) \cite{endf}.
The developed models were shown to have
a good performance in those works \cite{g4ce, g4dpn}
and experimental verification of the model
was also performed \cite{g4Kirams}.
To see the validity of the ENDF/B-VII.1 data for
$^{27}$Al(p,n)$^{27}$Si reaction,
we compare the results by the ENDF/B-VII.1 data
with the experimental cross section
data.
Other nuclear data such as
Japanese Evaluated Nuclear Data Library
High Energy File 2007 (JENDL/HE-2007) \cite{jendlHE},
and
TALYS-based evaluated nuclear data library (TENDL-2014) \cite{tendl}
can be available.
Results by these nuclear data are also compared to the experimental data
in this work.

$^{27}$Si with $T_{1/2}$ of 4.16 s emits electron-neutrinos
through radioactive decay processes
such as $\beta^{+}$ decay and electron capture (EC).
Radioactive decay processes for $^{27}$Si
and subsequent emitted electron-neutrino energy distributions are simulated
by using ``G4RadioactiveDecay" \cite{G4RDM_0, G4RDM_1} class
based on the Evaluated Nuclear Structure Data File (ENSDF) \cite{G4ensdf}.

\section{Results
\label{sec3}}

\subsection{Benchmarking simulations with experimental data
\label{sec3_1}}

\begin{figure}[tbp]
\begin{center}
\epsfig{file=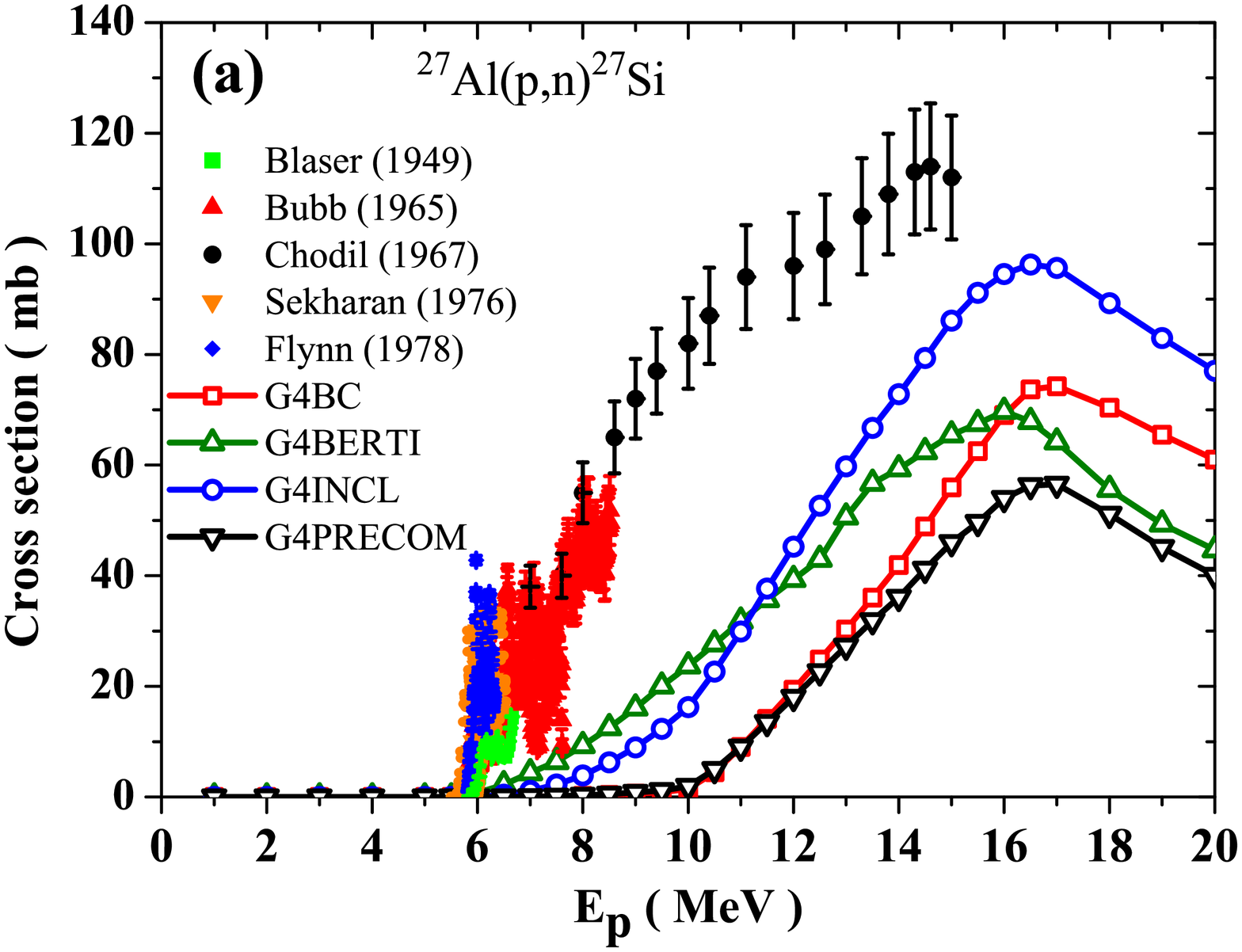, width=4.5in}
\epsfig{file=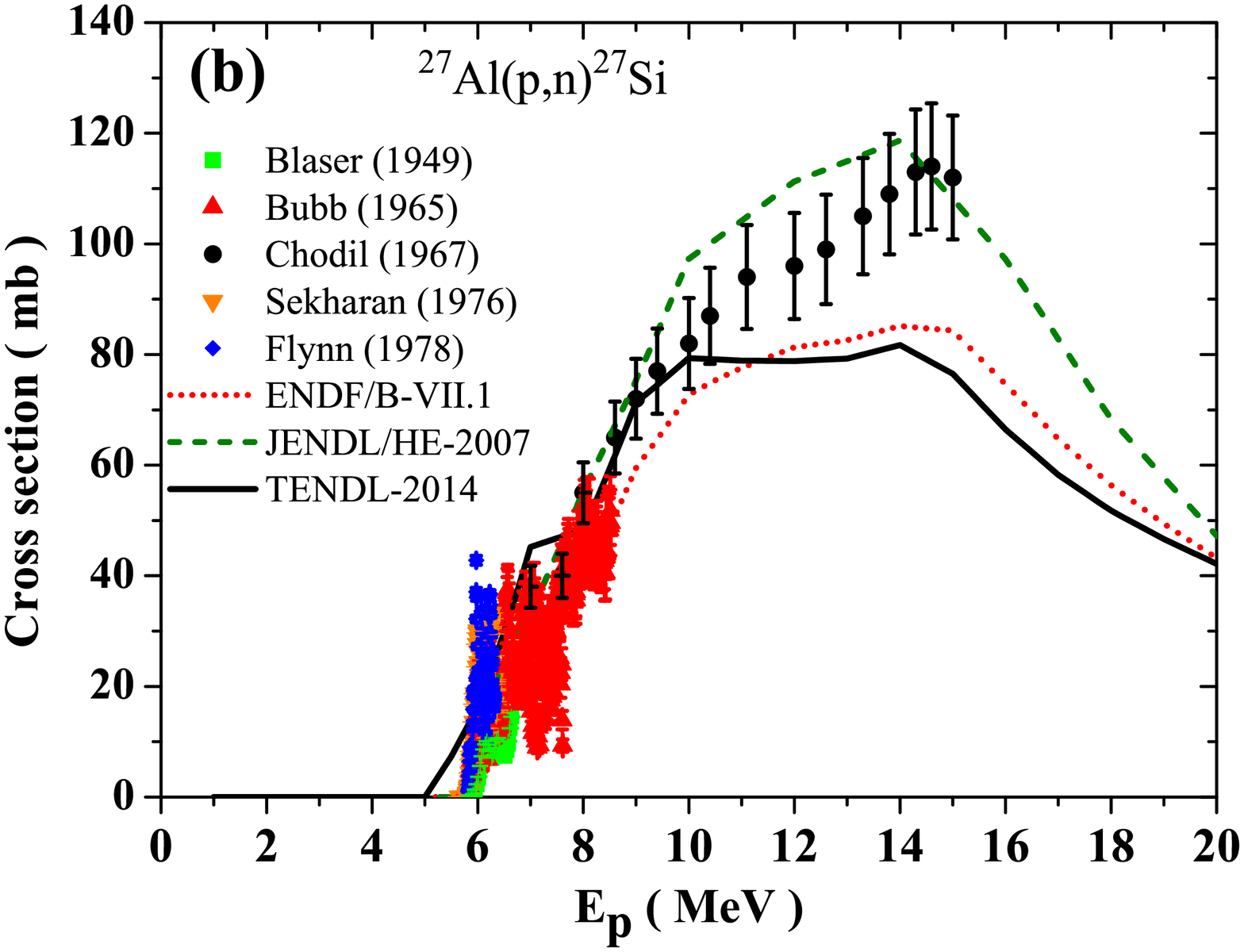, width=4.5in}
\end{center}
\caption{(Color online) (p,n) reaction cross-sections on the $^{27}$Al.
The symbols denote various experimental data taken from
the EXFOR database \cite{exfor}.
The experimental data are compared with the calculated cross sections
using hadronic models of GEANT4 and nuclear data model in panel (a) and (b),
respectively.}
\label{fig1}
\end{figure}

We first calculate proton induced $^{27}$Si production cross sections
from $^{27}$Al target by using the hadronic models of GEANT4
mentioned in Sec. \ref{sec2}.
The calculated results are compared with the experimental data
taken from the EXFOR database \cite{exfor}.
Results by the nuclear data are also compared with the measured data.

Cross sections for $^{27}$Al(p,n)$^{27}$Si reaction
with respect to the incident proton energy (E$_{p}$)
are plotted in Fig. \ref{fig1}.
The comparison of the calculated cross sections
obtained from hadronic models of GEANT4
with the experimental data
is given in Fig. \ref{fig1} (a), where none of the models can reproduce
the experimental cross sections
satisfactorily.
All the hadronic models considered in this work
underestimate the experimental cross sections
in $E_{p} \lesssim $ 15 MeV region.
Discrepancies between the calculated values and the measured those
increase as the incident proton energy decreases.
These discrepancies can be understood if we consider its widespread usage 
in many different scientific fields.
In general, hadronic models of GEANT4
have been developed and tested for wide use.
None of the models is specialized
for both particular reaction channels and low energy regions \cite{g4ce, g4dpn}.

Therefore, in Fig. \ref{fig1} (b), we show
the results by more realistic nuclear data and compare to
the experimental EXFOR data.
In the incident proton energy $E_{p} \lesssim $ 10 MeV region,
all the nuclear data reproduce well the experimental data with error bar.
For 10 MeV $\leq  E_{p} \leq $ 15 MeV,
ENDF/B-VII.1 (TENDL-2014) underestimates the experimental data 
about $\sim$ 19.3\% ($\sim$ 20.7\%),
and JENDL/HE-2007 overestimates the data about $\sim$ 9.2\%.
A comparison of Fig. \ref{fig1} (b) with Fig. \ref{fig1} (a)
demonstrates that the nuclear data model can give more reasonable results
than those by hadronic models of GEANT4 in the $E_{p} \leq $ 15 MeV region.
Especially JENDL/HE-2007 data reproduce well the experimental data
within the experimental error ($\sim$ 10\%). 
Additional benchmark simulations for
$^{27}$Al(p,xn) and $^{27}$Al(p,x$\alpha$) reactions
are also performed in this work.
The results are presented in Appendix \ref{app},
which shows a tendency similar to the results shown in Fig. \ref{fig1},
{\it i.e.} the hadronic models are inferior to the nuclear models at least
in the present simulation.

\subsection{$^{27}$Si yield calculation
\label{sec3_2}}

\begin{table}
\caption{Numbers of $^{27}$Si produced in $^{27}$Al target in the units of [1/incident proton] $\times$ 10$^{-4}$. }
\begin{tabular}{c|cc}
\hline
  Physics Model     & Number of $^{27}$Si produced in the target &\\
                    & [1/incident proton] $\times$ 10$^{-4}$ &\\ \hline
  JENDL/HE-2007     & 5.637 $\pm$ 0.289 &\\
  TENDL-2014        & 4.435 $\pm$ 0.215 &\\
  ENDF/B-VII.1      & 4.296 $\pm$ 0.175 &\\
  G4PRECOM          & 0.981 $\pm$ 0.032 &\\
\hline
\end{tabular}
\label{table_1}
\end{table}

\begin{figure}[tbp]
\begin{center}
\epsfig{file=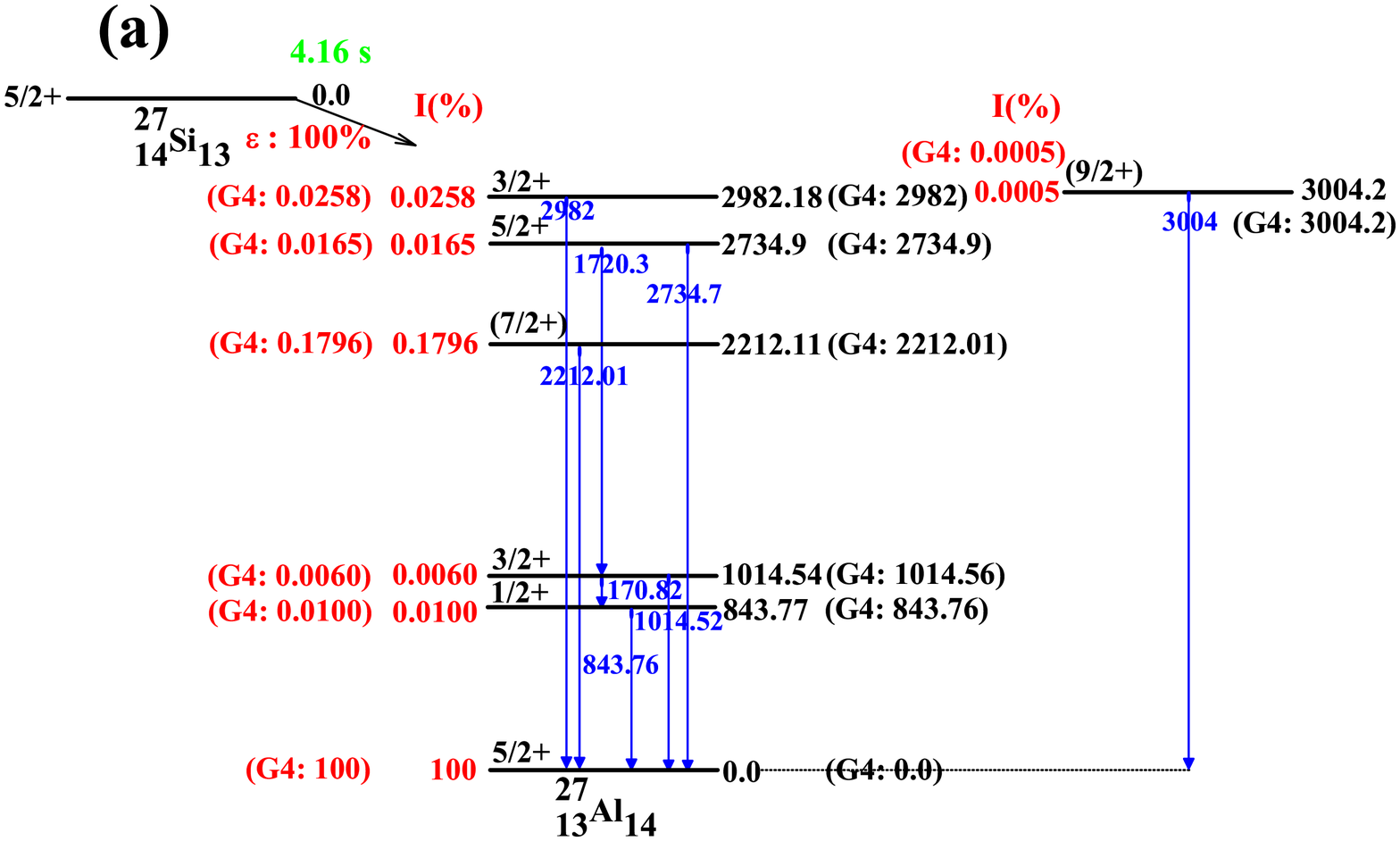, width=5in}
\epsfig{file=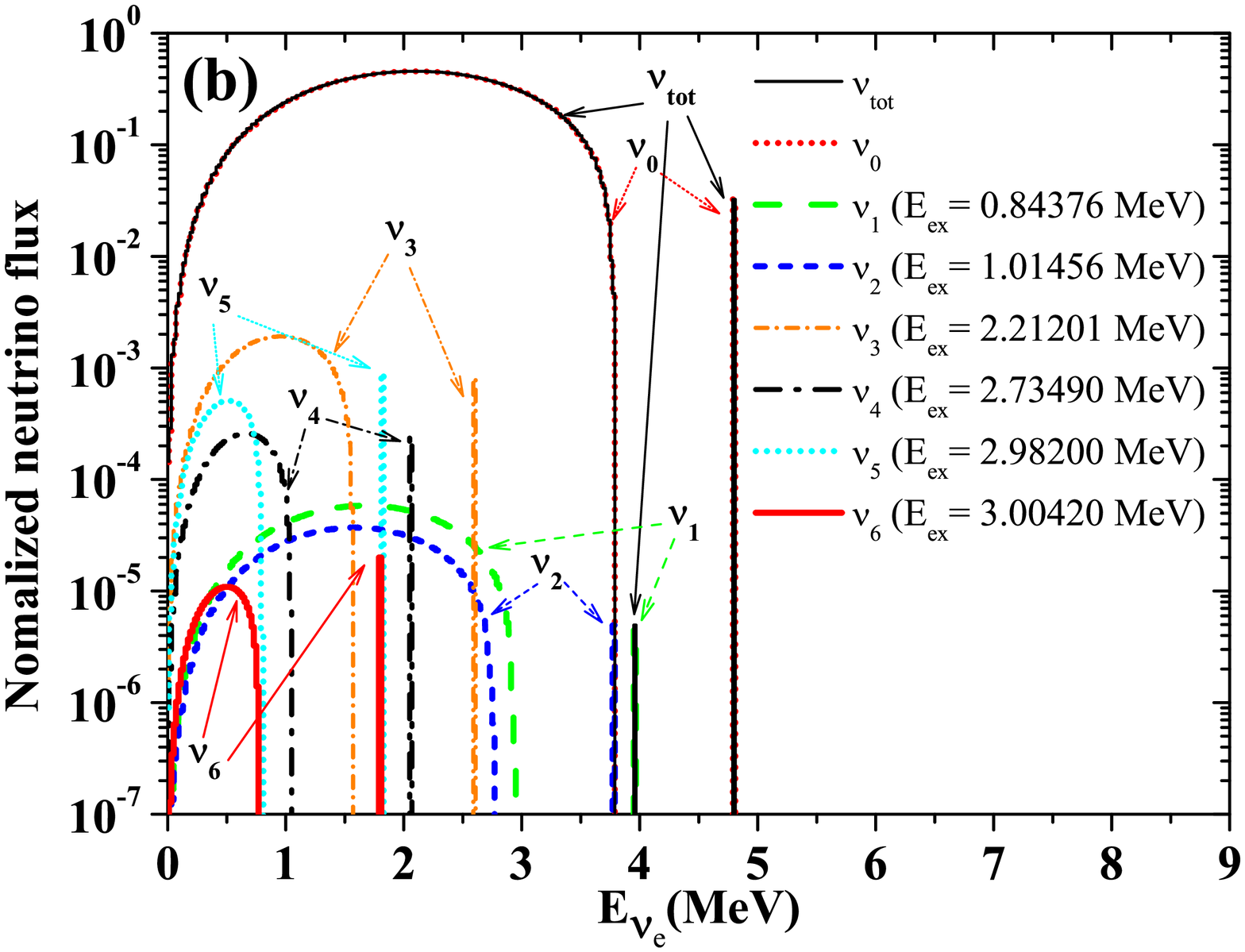, width=5in}
\end{center}
\caption{(Color online)
Upper panel (a) represents the energy levels (in the right hand side)
and their decay scheme with the branching ratio (BR) (in the left hand side) of $^{27}$Si.
Values in parenthesis with G4 denote those used in GEANT4.
Lower panel (b) notes the normalized energy distribution of
electron-neutrinos through the decay of $^{27}$Si.
E$_{ex}$ means the excitation level (MeV) of the daughter $^{27}$Al reached after the decay.
${\nu}_{0}$ means the electron-neutrinos with the residual $^{27}$Al in the ground state.
${\nu}_{1}$, ${\nu}_{2}$, ${\nu}_{3}$, ${\nu}_{4}$, ${\nu}_{5}$ and ${\nu}_{6}$
denote the electron-neutrinos leaving the $^{27}$Al with the excitation level
0.84376, 1.01456, 2.21201, 2.73490, 2.98200 and 3.00420 MeV, respectively.
${\nu}_{tot}$ is the sum from ${\nu}_{0}$ up to ${\nu}_{6}$.
It should be noted that the large difference between $\nu_0$
and others comes from the multiplied BRs in panel (a).
}
\label{fig2}
\end{figure}

Production of $^{27}$Si through p + $^{27}$Al reaction
is considered in this study.
Incident proton energy is chosen to be 15 MeV
because of the following reasons.
(1) There exist only limited numbers of experimental data in the EXFOR database.
Moreover, most of the data are centered in the energy region $E_{p} \lesssim $ 15 MeV.
For this reason,
benchmark test of subsection \ref{sec3_1} for
both the nuclear data and the hadronic models of GEANT4
is performed in this energy region.
(2) By choosing this energy,
we can suppress productions of unnecessary unstable isotopes
which can make neutrino or anti-neutrino backgrounds
through beta decays.
For example,
$^{23}$Mg, $^{25}$Al and $^{26}$Si also produce electron-neutrino through
beta decay processes.
However, these isotopes are not generated
in the present $E_{p} \lesssim $ 15 MeV energy region,
because threshold energies for $^{27}$Al(p,n+$\alpha$)$^{23}$Mg,
$^{27}$Al(p,t)$^{25}$Al and $^{27}$Al(p,2n)$^{26}$Si
reactions are
15.488 MeV, 16.536 MeV and 19.615 MeV, respectively.
When the 15 MeV proton beam is used,
$^{26}$Al can be generated through
$^{27}$Al(p,d)$^{26}$Al (E$_{th}$ = 11.237 MeV) or
$^{27}$Al(p,n+p)$^{26}$Al (E$_{th}$ = 13.545 MeV) reactions. However,
it has a very long life time of $\sim$ 10$^{5}$ y.
Also, $^{27}$Al(p,2p)$^{26}$Mg (E$_{th}$ = 8.580 MeV),
$^{27}$Al(p,p+$\alpha$)$^{23}$Na (E$_{th}$ = 10.468 MeV),
$^{27}$Al(p,$\alpha$)$^{24}$Mg and
$^{27}$Al(p,$\gamma$)$^{28}$Si reactions
are also possible,
but the isotopes produced through the reactions are stable.

To calculate the production rate of $^{27}$Si isotope,
we use JENDL/HE-2007 and TENDL-2014 data
as well as ENDF/B-VII.1 data for the simulations.
Extracted nuclear data are converted into the GEANT4 format
by using a TNudy (rooT NUclear Data librarY) program \cite{TNudyOld, TNudy}
which is a nuclear data manipulation package based on
the ROOT system \cite{Root}.
The program can easily search, access and visualize the nuclear data
relevant for user's specific purpose \cite{TNudyOld, TNudy}.
Simulations with G4PRECOM are also performed for the comparison.

The $^{27}$Al target is modeled as a cylinder of diameter 10 cm,
whose thickness is chosen as 0.1 cm.
The 1 cm thick $^{nat}$C layer is placed behind the target
for proton stopper.
Numbers of $^{27}$Si produced in $^{27}$Al target are tabulated
in Table \ref{table_1}.

\subsection{Energy spectra of electron-neutrinos from decay of $^{27}$Si
\label{sec3_3}}

Electron-neutrinos are emitted from $^{27}$Si
through ${\beta}^{+}$ decay and/or an electron capture (EC) process.
\begin{eqnarray}
{\rm {^{27}Si}} \to {\rm {^{27}Al}} + e^{+} + \nu_{e},
\label{eq:decay27Si_1}
\end{eqnarray}
\begin{eqnarray}
{\rm {^{27}Si}} + e^{-} \to {\rm {^{27}Al}} + \nu_{e}.
\label{eq:decay27Si_2}
\end{eqnarray}
For ${\beta}^{+}$ decay,
a bound proton of $^{27}$Si is changed into a neutron
and an electron-neutrino is emitted from the parent nucleus
with a positron in Eq. ~(\ref{eq:decay27Si_1}).
The neutrinos generated by ${\beta}^{+}$ decay
have continuous energy distributions.
As a different type of radioactive decay,
$^{27}$Si can also be converted into $^{27}$Al absorbing
an inner shell electron (e.g. K-, L-, M-shell electrons),
so-called EC or inverse ${\beta}^{+}$ decay in Eq. ~(\ref{eq:decay27Si_2}).
During the process,
the electron-neutrinos are produced
with discrete energy due to two-body kinematics.

Decay and level schemes of $^{27}$Si are drawn in Fig. \ref{fig2} (a).
Life time of $^{27}$Si, branching ratios
and excitation energies are based on
the ENSDF data.
The corresponding values used for GEANT4 simulations
are also noted in parentheses.
About 99.76\% of $^{27}$Si decay into a ground state of $^{27}$Al,
and the remaining $\sim$0.23\% of $^{27}$Si decay to
one of six different excited states of $^{27}$Al.
During the decay process,
$^{27}$Si emits electron-neutrinos
whose energies are restricted by each state of $^{27}$Al.

Figure \ref{fig2} (b) show the normalized 
neutrino energy spectra from decay of $^{27}$Si. 
In this work, we simulate the decay of 10$^{9} \times ^{27}$Si, 
but all results are divided by 10$^{9}$.
${\nu}_{0}$ means the electron-neutrinos
with the residual $^{27}$Al in the ground state.
${\nu}_{1}$, ${\nu}_{2}$, ${\nu}_{3}$, ${\nu}_{4}$, ${\nu}_{5}$ and ${\nu}_{6}$
denote the electron-neutrinos leaving the $^{27}$Al with the excitation level
0.84376, 1.01456, 2.21201, 2.73490, 2.98200 and 3.00420 MeV, respectively.
${\nu}_{tot}$ is the sum from ${\nu}_{0}$ up to
${\nu}_{6}$, where the ${\nu}_{0}$ contribution is dominant ($>$ 99.76\%),
while other contributions of ${\nu}_{1}$ $\sim$ ${\nu}_{6}$ are marginal
because of very small branching ratios.

In Fig. \ref{fig2} (b), maximum monoenergy neutrino of 4.813 MeV
in ${\nu}_{0}$ as well as other monoenergies stem
from the two body EC reaction, and
continuous energy spectra come from the three body ${\beta}^{+}$ decay.
Energy differences between the maximum energy of neutrino
generated by $\beta^{+}$ decay
and the discrete neutrino energy by EC are $\sim$ 1.02 MeV.
This feature can be understood by considering
a sum of electron and positron masses
in Eqs. ~(\ref{eq:decay27Si_1}) and ~(\ref{eq:decay27Si_2}).
Next monoenergy peak from ${\nu}_{1}$
appears at E$_{\nu}$ = 3.97 MeV in the figure,
but it's intensity is much less than ${\nu}_{0}$ due to EC
by a factor of $\sim$ 6500
because of the small branching ratio of the corresponding excited state shown
in the panel (a) in Fig. \ref{fig2}. 
There is also a monoenergy peak of ${\nu}_{2}$ at 3.799 MeV
in Fig. \ref{fig2} (b),
but whose intensity is also very small like ${\nu}_{1}$ from EC.
As the excitation energy increases,
neutrino energy spectra are shifted to lower energies.
These shifts of the spectra are consistent with the difference
between the excitation energies.

\section{Discussion for electron-neutrino detection
\label{sec4}}

In the previous sections,
we proposed an accelerator based new artificial
electron-neutrino source generated from decay of $^{27}$Si
for experiments of the vacuum-matter transition
region. As noted in section \ref{sec3_2},
we produce a specific unstable
isotope $^{27}$Si in our case by choosing the 15 MeV proton beam.
With the aid of the characteristics due to both
an accelerator-based neutrino and
a relatively short life time of $^{27}$Si isotope (a few seconds),
we can control an artificial neutrino beam,
and thus remove background neutrinos
for the detection
in the following way.
During the proton beam on $^{27}$Al target,
electron-neutrinos are generated from $^{27}$Si.
We can obtain signals (S$^{\prime}$ = S$\it{_{s.}}$ + S$\it{_{b.g.}}$) in detecting-materials
interacting with both the neutrinos from $^{27}$Si (S$\it{_{s.}}$)
and background neutrinos (S$\it{_{b.g.}}$).
When the beam is switched off, on the other hand,
the neutrinos are not produced.
We can thus only obtain signals (S = S$\it{_{b.g.}}$)
from the background.
By subtracting the S from the S$^{\prime}$
we obtain the signal S$\it{_{s.}}$.
Therefore, through the beam on and off,
we may remove background neutrinos such as solar neutrinos,
atmospheric neutrinos, geo-neutrinos, etc.

\begin{figure}[tbp]
\begin{center}
\epsfig{file=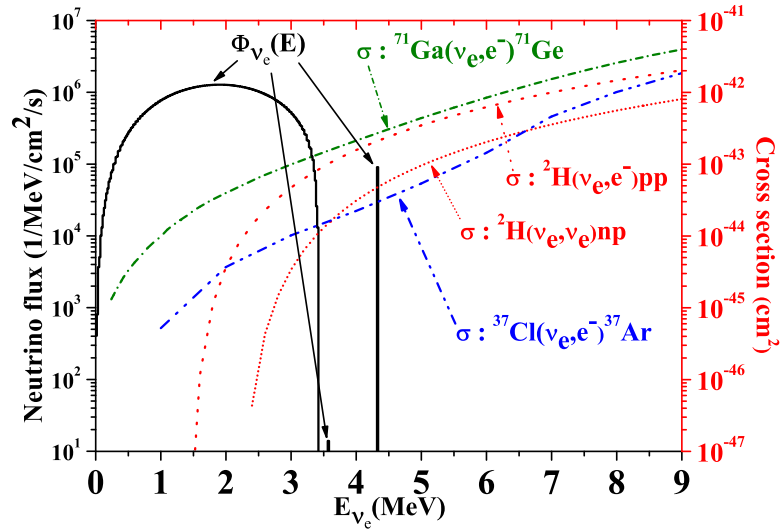, width=4.5in}
\end{center}
\caption{(Color online)
Generated neutrino spectrum. The black solid lines represent the expected electron-neutrino
energy spectrum at a distance of 10 m from the $^{27}$Al target.
The neutrinos are generated from decay of $^{27}$Si produced
by the 15 MeV and 10 mA proton beam on a $^{27}$Al target.
For further discussions, we show theoretical cross section for
$^{37}$Cl($\nu_{e}$,e$^{-}$)$^{37}$Ar \cite{nu_37Cl},
$^{71}$Ga($\nu_{e}$,e$^{-}$)$^{71}$Ge \cite{nu_71Ga},
$^{2}$H($\nu_{e}$,e$^{-}$)pp \cite{nu_EFT1} and
$^{2}$H($\nu_{e}$,$\nu_{e}$)np reactions \cite{nu_EFT1}.
}
\label{fig3}
\end{figure}

The expected electron-neutrino energy spectrum
at a distance of 10 m from the $^{27}$Al target is plotted in Fig. \ref{fig3}.
The neutrinos are generated from decay of $^{27}$Si
produced by the 15 MeV and 10 mA proton beam on a $^{27}$Al target.
Although we assume the current of proton beam to 10 mA in this work,
higher current beams from future's high power proton facilities
can be available as reported in Refs. \cite{highACC_2, highACC_3}.
The neutrino flux $\Phi_{\nu_{e}}$ is
evaluated by using GEANT4 with JENDL/HE-2007 data
and G4RadioactiveDecay.
Main contribution of the spectrum comes from the ${\nu}_{0}$
through $\beta^{+}$ decay
and the highest energy of $\sim$ 4.813 MeV
is emitted from the ${\nu}_{0}$ by the EC,
as explained in Fig. \ref{fig2}.

\begin{figure}[tbp]
\begin{center}
\epsfig{file=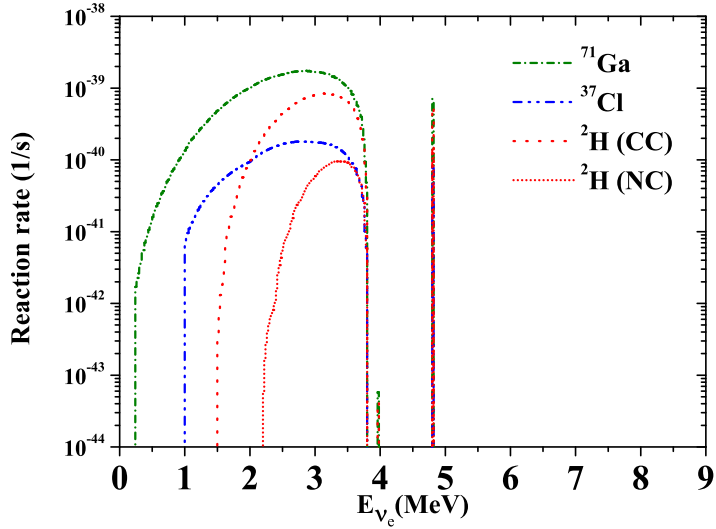, width=4.5in}
\end{center}
\caption{(Color online)
Electron-neutrino
induced reaction rates via
$^{37}$Cl($\nu_{e}$,e$^{-}$)$^{37}$Ar,
$^{71}$Ga($\nu_{e}$,e$^{-}$)$^{71}$Ge,
$^{2}$H($\nu_{e}$,e$^{-}$)pp and
$^{2}$H($\nu_{e}$,$\nu_{e}$)np reactions.
The neutrinos are generated from decay of $^{27}$Si which are
produced by the 15 MeV and 10 mA proton beam on a $^{27}$Al target.
Results of the cross sections in Fig. \ref{fig3} are exploited for this reaction rate calculation.
}
\label{fig4}
\end{figure}

\subsection{Radiochemical detectors
\label{sec4_1}}
Here we discuss two different kinds of detections for the neutrino generated
in the present simulation.
One is the radiochemical detection technique,
which has been widely used for the solar neutrino detection experiments.
The cross sections for
$^{37}$Cl($\nu_{e}$,e$^{-}$)$^{37}$Ar \cite{nu_37Cl},
$^{71}$Ga($\nu_{e}$,e$^{-}$)$^{71}$Ge \cite{nu_71Ga},
$^{2}$H($\nu_{e}$,e$^{-}$)pp \cite{nu_EFT1} and
$^{2}$H($\nu_{e}$,$\nu_{e}$)np \cite{nu_EFT1} reactions
are plotted
in Fig. 3 for the radiochemical detection
of the neutrino from $^{27}$Si.

The accelerator neutrino can be captured via the weak charged current reaction
in the target nucleus such as
$^{37}$Cl and $^{71}$Ga, and then new radioactive isotopes are produced.
Those isotopes are chemically extracted and counted using their decay.
Therefore, the neutrino source suggested in this work
can be used for the calibration both $^{37}$Cl and $^{71}$Ga solar neutrino
detector system.

\begin{table}
\caption{
Radiochemical neutrino detector materials.
The total reaction rates are calculated by both
the electron-neutrino spectra obtained from this work
and the cross sections mentioned in Fig. \ref{fig3}.
The peak-to-total ratio in the 4th column
means the ratio of the contribution from the ${\nu}_{0}$ due to EC reaction
to the total reaction rate.
}
\begin{tabular}{c|c|c|cc}
\hline
    Isotope    & Reaction                                           & Total reaction rate (1/s) & Peak-to-total ratio (\%) &\\ \hline
    $^{2}$H    &  $\nu_{e}$ + $^{2}$H   $\to$ $e^{-}$ + p + p       & 4.888E-38               & 1.127                 &\\
               &  $\nu_{e}$ + $^{2}$H   $\to$ $\nu_{e}$ + p + n     & 3.841E-39               & 3.790                 &\\
    $^{37}$Cl  &  $\nu_{e}$ + $^{37}$Cl $\to$ $e^{-}$ + $^{37}$Ar   & 1.496E-38               & 0.579                 &\\
    $^{71}$Ga  &  $\nu_{e}$ + $^{71}$Ga $\to$ $e^{-}$ + $^{71}$Ge   & 1.473E-37               & 0.478                 &\\
    \hline
\end{tabular}
\label{table_2}
\end{table}

Figure \ref{fig4} shows the reaction rates
calculated from the cross section in Fig. \ref{fig3} with respect to the neutrino energy
generated for a given condition.
For both $^{37}$Cl and $^{71}$Ga,
neutrinos having
the continuous energy (E$_{th} <$ E$_{\nu_{e}}$ $< \sim$ 3.79 MeV)
and monoenergies (3.97 MeV and 4.813 MeV)
interact with the isotopes.
Contributions of the neutrino with two monoenergies of 3.97 MeV and 4.813 MeV
are marginal.
Their fractions in the total reaction rate of $^{71}$Ga ($^{37}$Cl)
turns out to be only 0.00004\% (0.00004\%) and 0.47809\% (0.57929\%)
for 3.97 MeV and 4.813 MeV, respectively.
Consequently,
we can obtain the reaction rates or flux-averaged cross sections
for both $^{37}$Cl($\nu_{e}$,e$^{-}$)$^{37}$Ar and
$^{71}$Ga($\nu_{e}$,e$^{-}$)$^{71}$Ge reactions
in the neutrino energy region less than $\sim$ 3.79 MeV.
This energy region is close to the vacuum-matter transition region
in the solar neutrino physics.

One can see that widths of energy distributions of
reaction rates for deuteron targets are more
narrow than those for both $^{37}$Cl and $^{71}$Ga
because of the reaction thresholds.
With this feature,
we can also have a chance to obtain reaction rates or
energy averaged cross sections for
$^{2}$H($\nu_{e}$,e$^{-}$)pp and
$^{2}$H($\nu_{e}$,$\nu_{e}$)np reactions
with narrow neutrino energy region.
Total reaction rates and the peak-to-ratio
for reactions of the results in Fig. \ref{fig4}
are tabulated in Table \ref{table_2}.

As another candidate of the radiochemical neutrino detection material,
$^{127}$I was also suggested in Ref. \cite{127I_1}.
A first cross section measurement
for $^{127}$I($\nu_{e}$,$e^{-}$)$^{127}$Xe reaction
was done by using the $\nu$ flux from the decay of stopped muons
at the Los Alamos Meson Physics Facility (LAMPF) \cite{127I_exp}.
Their results are useful for super novae neutrino measurements
because of the relatively high energy region of the neutrinos from LAMPF.
But neutrino sources suggested in this work
which have smaller energy range ($\sim <$ 3.79 MeV) than LAMPF
can play
an important role of improving the solar neutrino detection
using $^{127}$I.

\begin{figure}[tbp]
\begin{center}
\epsfig{file=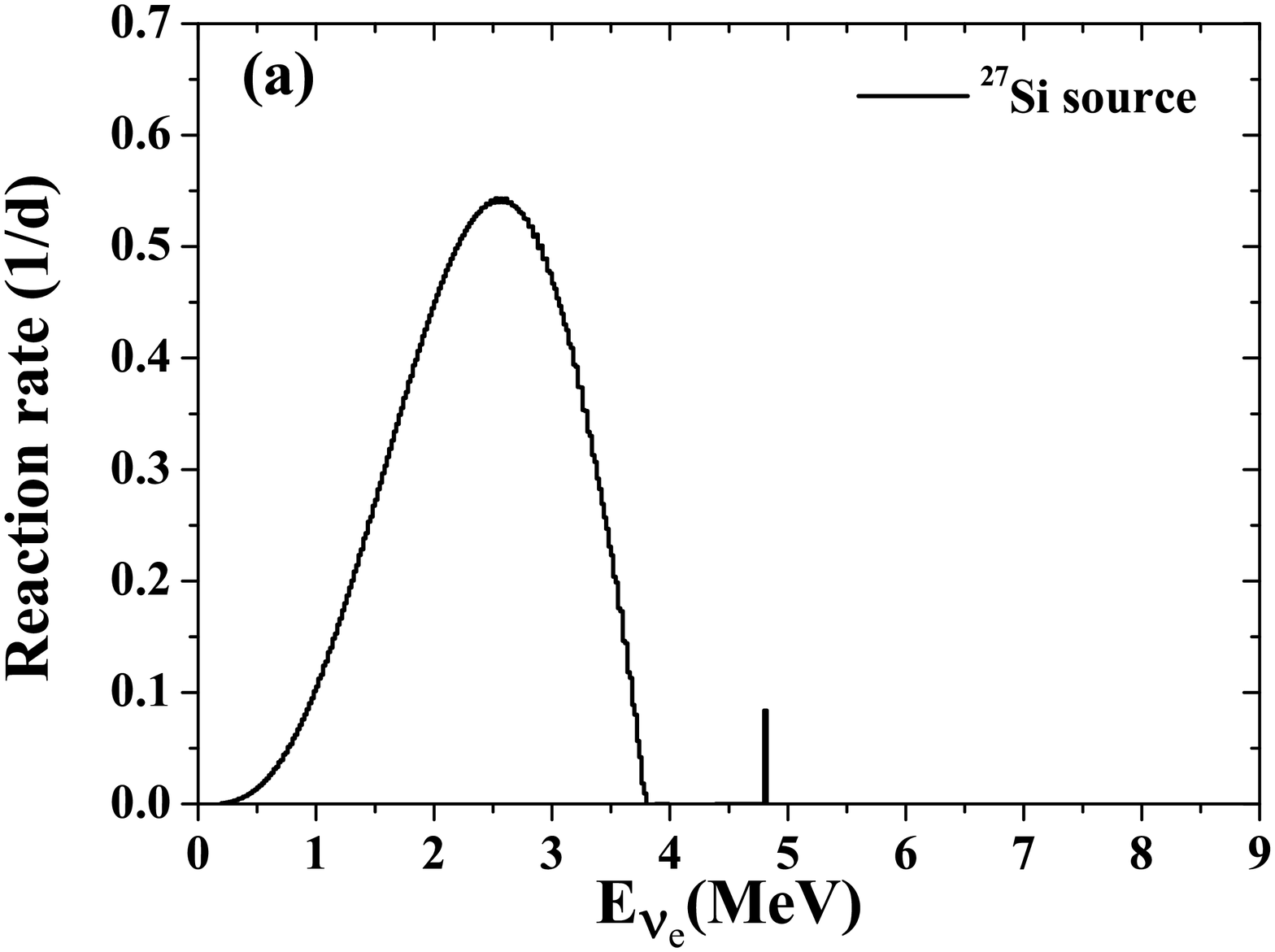, width=4.5in}
\epsfig{file=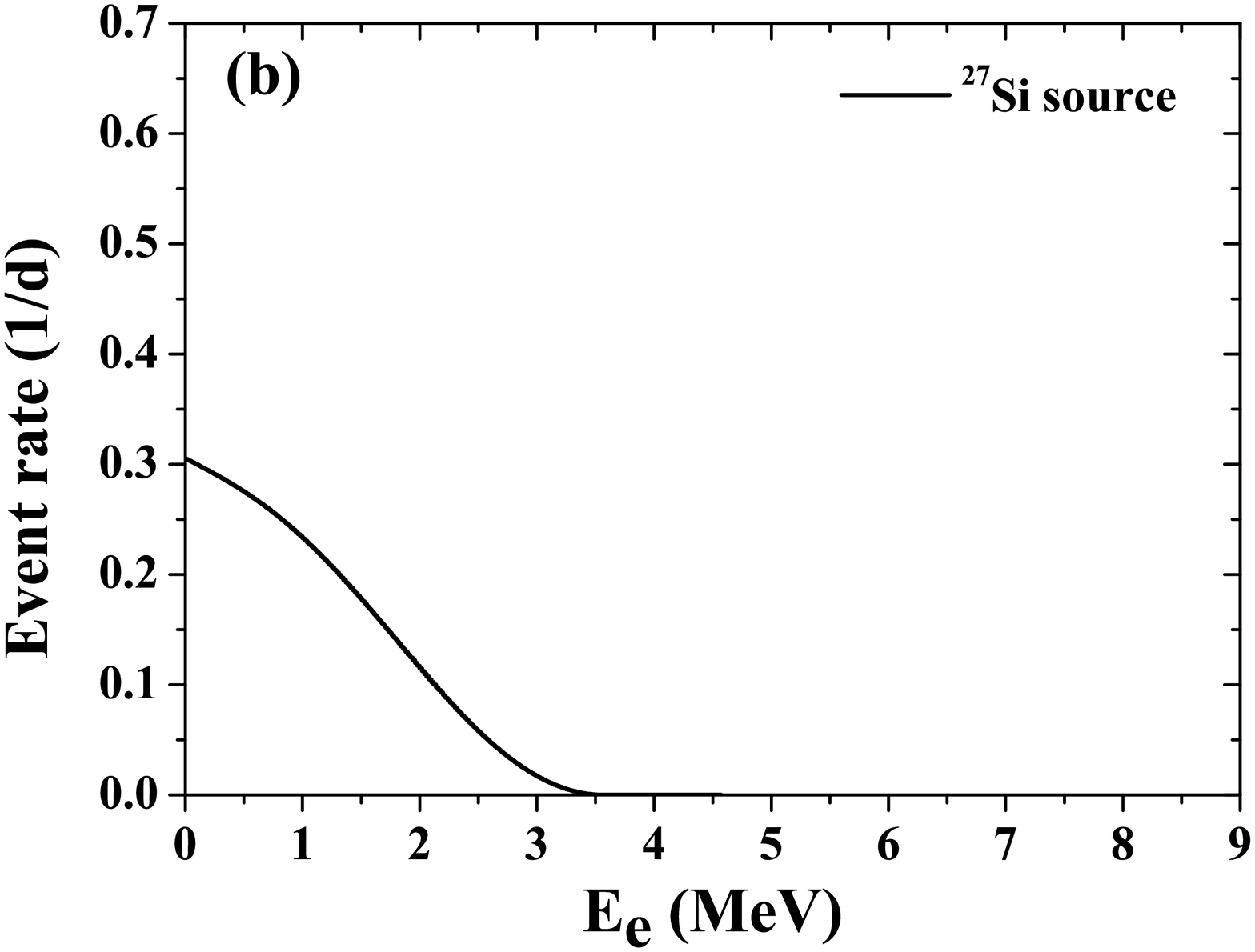, width=4.5in}
\end{center}
\caption{(Color online)
Panel (a) shows the reaction rate for electron-neutrino electron ($\nu_{e}$-e$^{-}$)
elastic scattering (ES) in the LENA detector.
The neutrinos are generated from decay of $^{27}$Si produced by the 15 MeV and 10 mA proton beams on a $^{27}$Al target.
The distance between the $^{27}$Si source
and the center of the LENA detector is assumed as 10 m.
Panel (b) presents recoil energy distribution of the outgoing electron
for the same condition.
}
\label{fig5}
\end{figure}

\subsection{Scintillator detectors
\label{sec4_2}}
In addition, the neutrino source from this work
can be useful for electron-neutrino detection studies
with future's gigantic liquid-scintillator detectors (LSDs)
such as LENA \cite{LENA_0}.
Through electron-neutrino electron ($\nu_{e}$-e$^{-}$) elastic scattering (ES),
neutrinos can be indirectly detected by the outgoing scattered electron, which electron can be identified by means of the scintillation light
produced in the liquid scintillator.
As a part of the European Large Apparatus for Grand Unification
and Neutrino Astrophysics (LAGUNA) design study,
the $\sim$ 50 kt LSD LENA (Low Energy Neutrino Astronomy) would
have the feature of low detection threshold, good energy resolution,
particle identification with efficient background discrimination
\cite{LENA_a1, LENA_a2, LENA_a3}.

With the electron-neutrino flux $\Phi_{\nu_{e}} (E_{\nu})$ in Fig. \ref{fig3},
we obtain the expected reaction rate ($R_{\nu_{e}}$) for LENA LSD.
Differential reaction rate for the neutrinos from $^{27}$Si
can be obtained by integrating over the neutrino energy \cite{LENA_ER}
\begin{eqnarray}
\frac{d R_{\nu_{e}}}{dT}
= n_{e} \int^{E_{max}}_{0} dE_{\nu} \Phi_{\nu_{e}} (E_{\nu}) {\rm{P}}_{ee}(E_{\nu}) \frac{d \sigma}{dT}.
\label{eq:LENA_1}
\end{eqnarray}
where $T$ is the kinetic energy of the recoil electron,
$n_{e}$ is the number of the electrons within the fiducial volume of LENA,
$\Phi_{\nu_{e}} (E_{\nu})$ is the electron-neutrino flux,
$E_{max}$ is the maximal neutrino energy,
${\rm{P}}_{ee}(E_{\nu})$ is the energy dependent electron-neutrino survival probability,
and $E_{\nu}$ is the energy of the incident neutrino.
The $\frac{d \sigma}{dT}$ in Eq. ~(\ref{eq:LENA_1})
is the
differential cross sections for neutrino-electron scattering
which is give as \cite{nu_eElas}
\begin{eqnarray}
\frac{d \sigma}{dT}(\nu_{l}e \to \nu_{l}e)
= \frac{2G^{2}_{\mu} m_{e}}{\pi E^{2}_{\nu} } [a^{2}E^{2}_{\nu} + b^{2}(E_{\nu} - T)^{2} - ab m_{e} T ],
\label{eq:LENA_2}
\end{eqnarray}
where $G_{\mu}$ is the Fermi constant,
$m_{e}$ is the mass of electron,
and $a = - \frac{1}{2} - s^{2}$ and $b = - s^{2}$.
The $s^{2} = $ sin$^{2}$ $\theta_{W}$
where $\theta_{W}$ is weak mixing angle.
In our case,
we use ${\rm{P}}_{ee}(E_{\nu})$
given by \cite{PeeRef1}
\begin{eqnarray}
P({\nu}_{e} {\to} {\nu}_{e})
= 1 - {\rm{sin}}^{2} 2 \theta_{13} S_{23} - c^{4}_{13}{\rm{sin}}^{2} 2 \theta_{12} S_{12},
\label{eq:Pee_1}
\end{eqnarray}
where $S_{23} = {\rm{sin}}^{2}(\Delta m^{2}_{32} L / 4 E)$ and
$S_{12} = {\rm{sin}}^{2}(\Delta m^{2}_{21} L / 4 E)$.

Expected reaction rate for $\nu_{e}$-e$^{-}$ ES
is plotted in Fig. \ref{fig5} (a).
One can notice that
dominant contribution of the reaction rate is
the electron-neutrino
with the energy range E$_{th} <$ E$_{\nu_{e}}$ $< \sim$ 3.79 MeV
and sub-dominant contribution is the neutrino with a monoenergy 4.813 MeV.
This feature can give us a chance to study
for solar neutrino detections via $\nu_{e}$-e$^{-}$ ES
in the transition region between vacuum-dominated and matter-enhanced,
where experimental understanding is still insufficient.
Also, this may provide a test ground for
the consistency of the standard model (SM),
the determination of precision electroweak parameters, etc \cite{nu_eElas}.
Total expected reaction rate is evaluated as 51.5392 count per day (cpd),
and expected energy spectrum of recoil electron (event rate) via $\nu_{e}$-e$^{-}$ ES
is also plotted in Fig. \ref{fig5} (b).

\begin{figure}[tbp]
\begin{center}
\epsfig{file=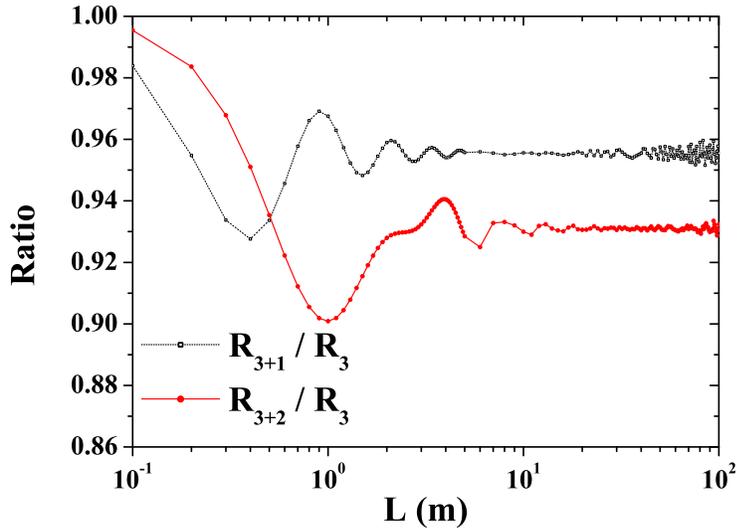, width=4.5in}
\end{center}
\caption{(Color online)
The ratio of the total reaction rate with P$_{3+1}$ (or P$_{3+2}$) to
the reaction rate with P$_{3}$ which is the same as P in Eq. ~(\ref{eq:Pee_1}).
L means the distance between the $^{27}$Si source
and the center of the LENA detector.
}
\label{fig6}
\end{figure}

\begin{figure}[tbp]
\begin{center}
\epsfig{file=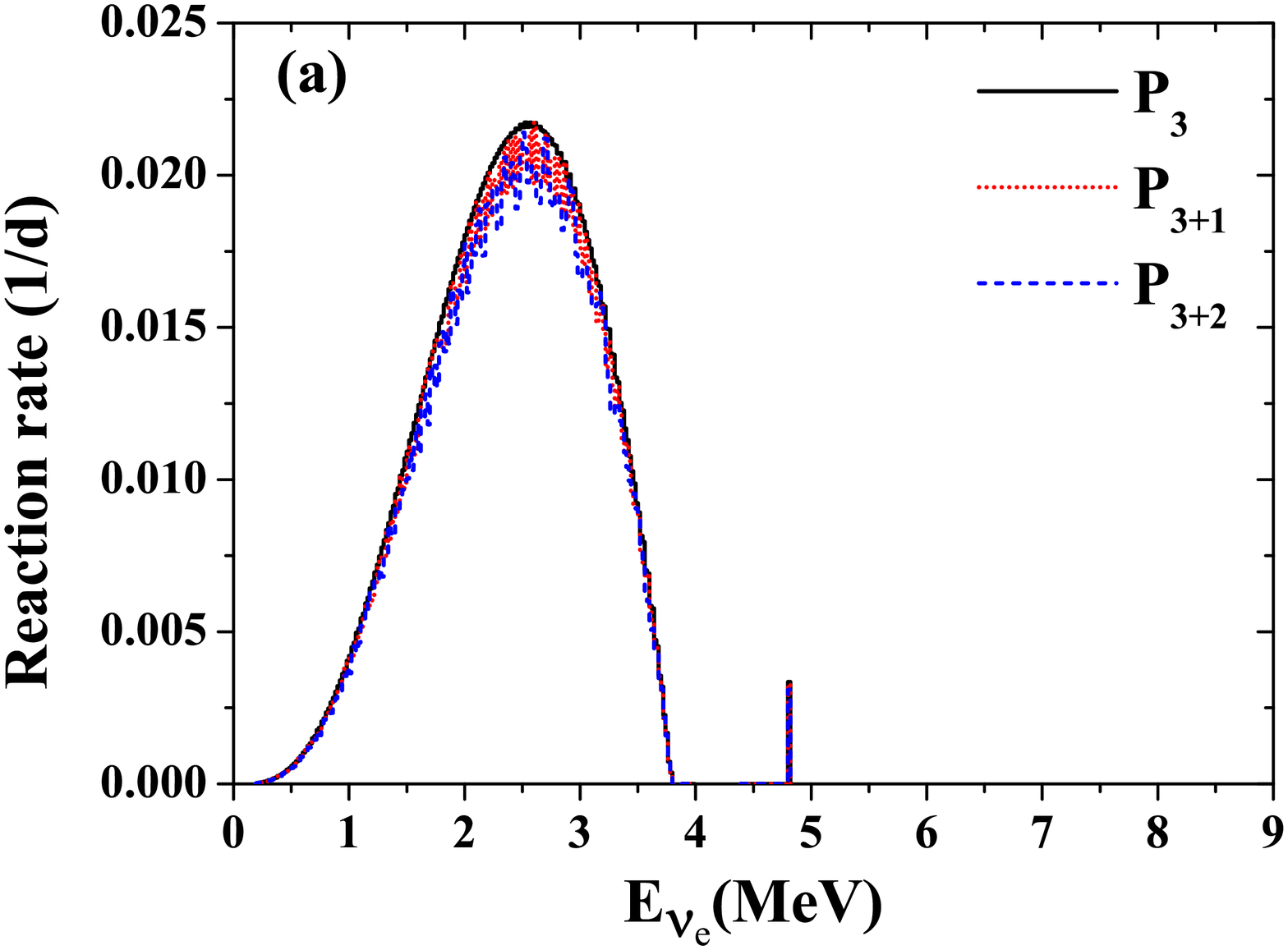, width=5in}
\epsfig{file=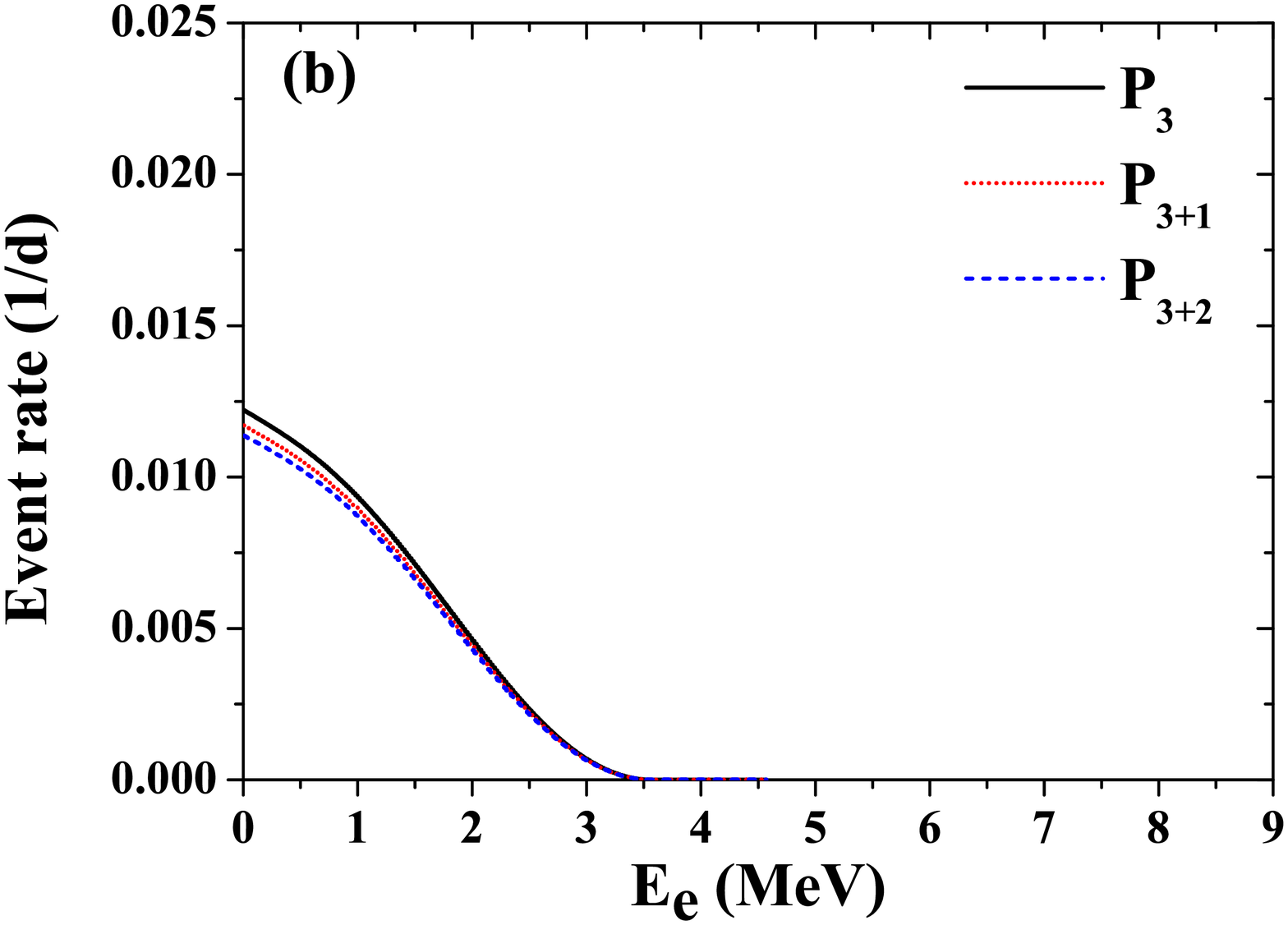, width=5in}
\end{center}
\caption{(Color online) Effects of the sterile neutrinos on the reaction
rate of $\nu_{e}$-e$^{-}$ ES scattering,
and recoiled energy distribution of the outgoing electron.
Others are the same as Fig. \ref{fig5}.
P$_{3}$, P$_{3+1}$ and P$_{3+2}$ are
electron-neutrino survival probabilities in Eqs.
(\ref{eq:Pee_1}), (\ref{eq:P_31}) and (\ref{eq:P_32}), respectively.
}
\label{fig7}
\end{figure}

\section{Effects of possible sterile neutrinos
\label{sec5}}

By using our neutrino source,
we can also study the possibility
of the existence of fourth neutrino, sterile neutrino.
Electron-neutrino survival probabilities
in the 3+1 and 3+2 scenarios can be written as \cite{sterileNu1}
\begin{eqnarray}
P_{\rm{3+1}}
= 1 - 4|U_{e4}|^{2}(1-|U_{e4}|^{2}){\rm{sin}}^{2}(\Delta m^{2}_{41} L / E),
\label{eq:P_31}
\end{eqnarray}
\begin{eqnarray}
P_{\rm{3+2}} =
1 &-& 4[ (1-|U_{e4}|^{2}-|U_{e5}|^{2}) \nonumber \\
&\times& (|U_{e4}|^{2}{\rm{sin}}^{2}(\Delta m^{2}_{41} L / E) \nonumber \\
&+& |U_{e5}|^{2} {\rm{sin}}^{2}(\Delta m^{2}_{51} L / E)) \nonumber \\
&+& |U_{e4}|^{2} |U_{e5}|^{2} {\rm{sin}}^{2}(\Delta m^{2}_{54} L / E)].
\label{eq:P_32}
\end{eqnarray}
The parameters in Eqs. ~(\ref{eq:P_31}) and ~(\ref{eq:P_32})
are taken from Ref. \cite{sterileNu2}.

Figure \ref{fig6} shows the ratio of
the total reaction rate with P$_{3+1}$ (or P$_{3+2}$) to
the reaction rate with P$_{3}$ which is the same as P in Eq. ~(\ref{eq:Pee_1}).
L is the distance between the $^{27}$Si source
and the center of the LENA detector.
For L of a few ten's meters region,
the value of R$_{3+2}$/R$_{3}$ is about 0.93.
When L is chosen to 50 m,
expected R$_{3}$, R$_{3+1}$ and R$_{3+2}$ are
2.06, 1.97 and 1.92 cpd, respectively.
The R$_{3+2}$-to-R$_{3}$ ratio is 0.931,
which is very close to the value of
reactor experiments measured / expected rate = 0.927$\pm$0.023,
so-called the short baseline reactor antineutrino
anomaly \cite{reactorAno_1, reactorAno_2}.
Therefore, together with
an electron antineutrino disappearance search \cite{sterileNu1},
our electron-neutrino source can be used for
a precise measurement of the weak mixing angle
at $\Delta m^{2}$ $\sim$ 1 eV$^{2}$
relevant to the sterile neutrino.

Figure \ref{fig7} (a) and (b) show possible effects
by the sterile neutrinos on the expected reaction rate for $\nu_{e}$-e$^{-}$ ES
and the recoil electron energy spectra, respectively.
The neutrino source and their physical conditions
are the same as Fig. \ref{fig5}, but L is taken as 50 m in Fig. \ref{fig7}.
The results of Fig. \ref{fig7} are calculated with
the P$_{3}$, P$_{3+1}$ and P$_{3+2}$ models, where P$_{3}$, P$_{3+1}$ and P$_{3+2}$ are
electron-neutrino survival probabilities in Eqs.
(\ref{eq:Pee_1}), (\ref{eq:P_31}) and (\ref{eq:P_32}), respectively.

As shown, in Figs. \ref{fig6} and \ref{fig7},
sterile neutrinos would affect the reaction (event) rate,
Eq. ~(\ref{eq:LENA_1}), 10\% maximally, compared to the three active neutrino model.
In particular, the 3+2 model influences the reaction much more than the 3+1 model.
This feature would be a very interesting
if the neutrino source is realized with reasonable detectors.
Our suggestion may complement the simulation results
regarding the effect of the sterile neutrinos on the disappearance of
the antineutrino from the artificial $^{8}$Li source
in Ref. \cite{sterileNu1}.

\section{Summary
\label{sec6}}

In this work,
an artificial electron-neutrino source
through $^{27}$Al(p,n)$^{27}$Si reaction is investigated.
The unstable isotope, $^{27}$Si, produced by the reaction emits the electron-neutrinos via
$\beta^{+}$ decay or electron capture (EC) process,
whose energy regions are very close to the vacuum-matter transition region
in the solar neutrino physics.

After extensive comparative study of the $^{27}$Al(p,n)$^{27}$Si reaction 
with many nuclear data models, the JENDL/HE-2007 data turned out to
reproduce well the experimental data
within the experimental error ($\sim$ 10\%),
and thus $^{27}$Si yields are evaluated by using GEANT4
with JENDL/HE-2007 data.
Energy distributions of the electron-neutrino from $^{27}$Si
are obtained by using ``G4RadioactiveDecay" class based on
the Evaluated Nuclear Structure Data File (ENSDF).
Neutrinos produced by the simulation are shown 
to have continuous energy spectra ($<$ $\sim$ 3.79 MeV)
due to ${\beta}^{+}$ decay and discrete energy (4.81 MeV) caused by EC.

By using the electron-neutrino source confined 
in the neutrino energy region less than $\sim$ 3.79 MeV,
we presented the reaction rates or flux-averaged cross sections
for $^{37}$Cl($\nu_{e}$,e$^{-}$)$^{37}$Ar and
$^{71}$Ga($\nu_{e}$,e$^{-}$)$^{71}$Ge
as well as the deuteron target.
The neutrino energy region produced in this work is located
in the vacuum-matter transition region
in solar neutrino physics, where experimental
data for enough understanding are still insufficient.

In addition, we addressed that neutrino sources from this work
can also be used for electron-neutrino oscillation studies
with future's gigantic liquid-scintillator detectors (LSDs)
such as LENA.
Possibility of detecting the sterile neutrino
by using the electron-neutrino source and the LENA type detector
is also discussed in detail.
Our results are shown to enable us to properly disentangle
the sterile neutrino mixing from the present electron-neutrino source.

Finally, we summarize detailed expected numbers of physical quantities 
for the neutrino beam and the detection system suggested in this work. 
If we scatter the proton accelerator beam of 15 MeV and 10 mA 
off the $^{27}$Al target, we obtain $\sim$ 3.5$\times$10$^{13}$ $^{27}$Si isotope per second 
according to JENDL/HE-2007 data (see Table \ref{table_1}). 
Numbers of neutrinos expected to be emitted from the $^{27}$Si source 
are 1.28 $\times$ 10$^6$ per second if we detect them at 10 m (see Fig. \ref{fig3}). 
This number comes from the maximum value of the flux in Fig. \ref{fig2}, 0.46, 
around 2 MeV region by multiplying the $^{27}$Si number and dividing 4$\pi r^2$ with r = 10 m.

For the detection system, first we consider the radiochemical detector. 
We presented expected reaction rates for a nucleus target, 
such as $^2$H, $^{37}$Cl, and $^{71}$Ga (see Fig. \ref{fig4} and Table \ref{table_2}). 
If we multiply numbers of target nucleus, 
we can get more realistic reaction rates for each target system. 
In the case of scintillator target, such as LENA, assumed to be located at 10 m (50 m), 
we expect 51.53 (2.06) reactions of the $\nu_{e} - e$ elastic scattering 
(see Figs. \ref{fig5} (a) and \ref{fig7} (a)), 
respectively, if we run the artificial neutrino accelerator for a day. 
Also expected numbers of the event of the scattered electron 
are reported in Figs. \ref{fig5} (b) and \ref{fig7} (b). 
In addition, effects of the presumed sterile neutrinos $P_{3+1}$ and $P_{3+2}$ model 
are evaluated for the numbers of the events, 
which are shown to be less than 10\%, maximally, of that by $P_{3+0} = P$ model.

\section*{Acknowledgments}
The work of J. W. Shin is supported by
the National Research Foundation of Korea \ (Grant No. NRF-2015R1C1A1A01054083),
the work of M.-K. Cheoun is supported by
the National Research Foundation of Korea \ (Grant No. NRF-2014R1A2A2A05003548).

\appendix
\section{Additional benchmark test for proton on $^{27}$Al
\label{app}}

\begin{figure}[tbp]
\epsfig{file=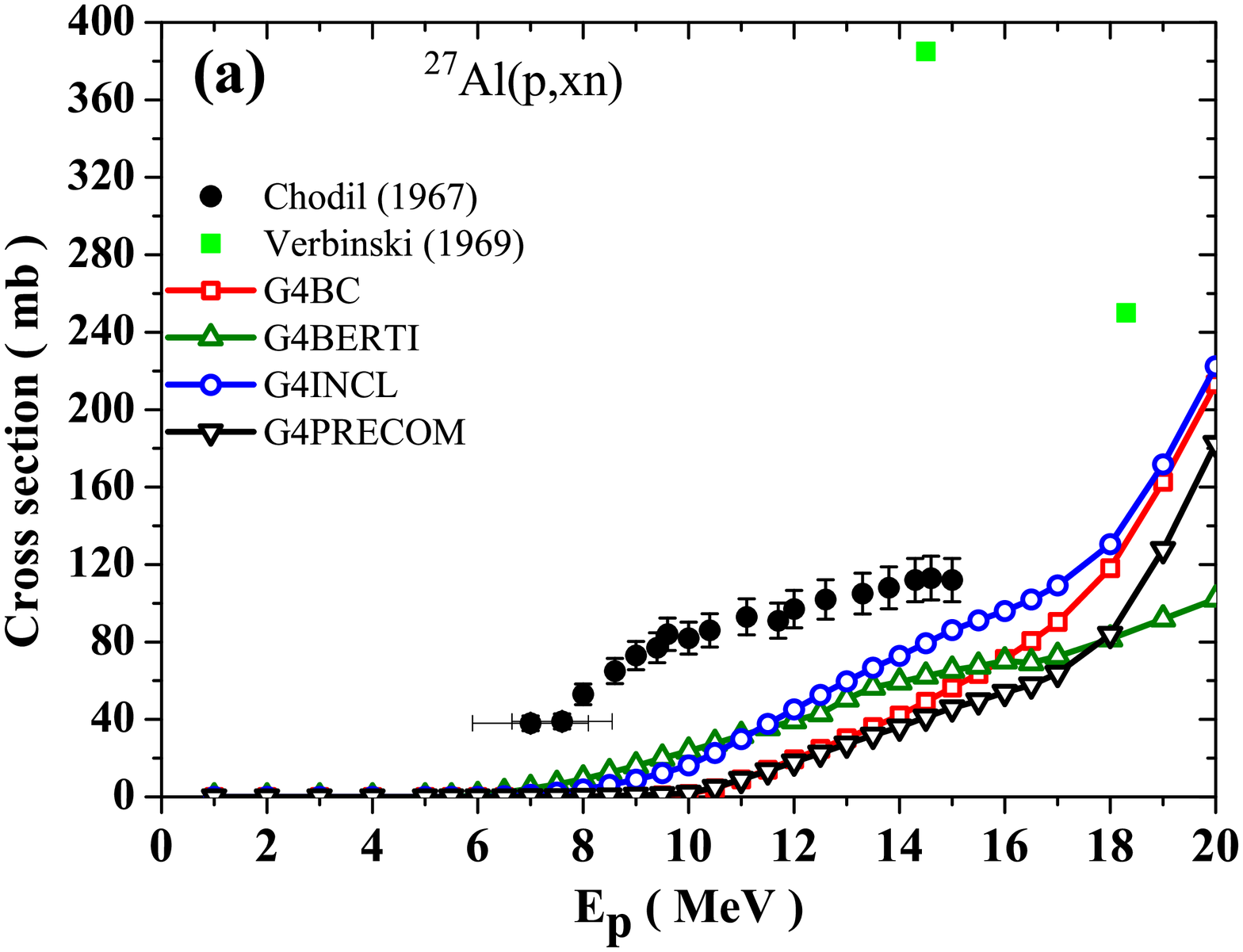, width=2.7in}
\epsfig{file=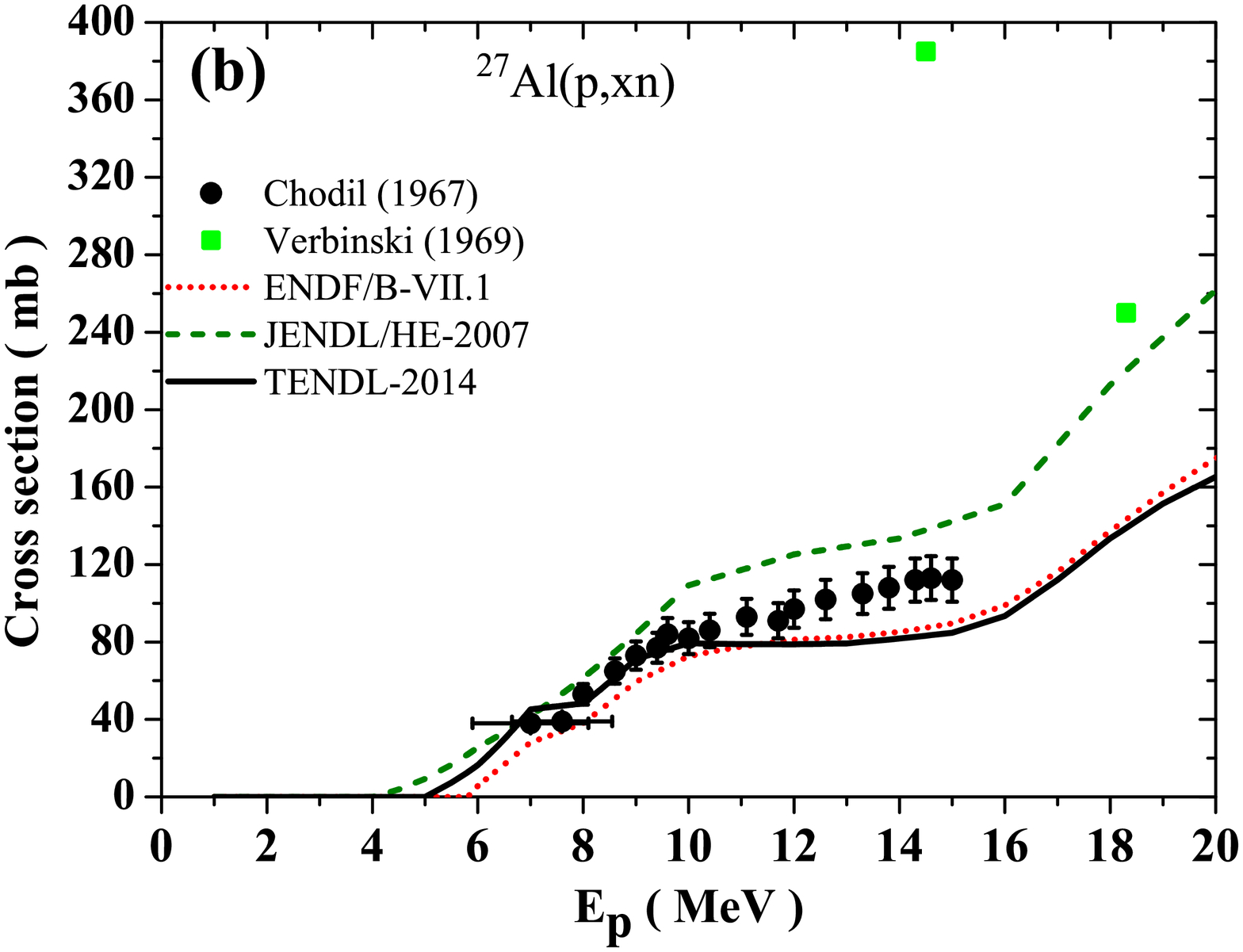, width=2.7in}
\epsfig{file=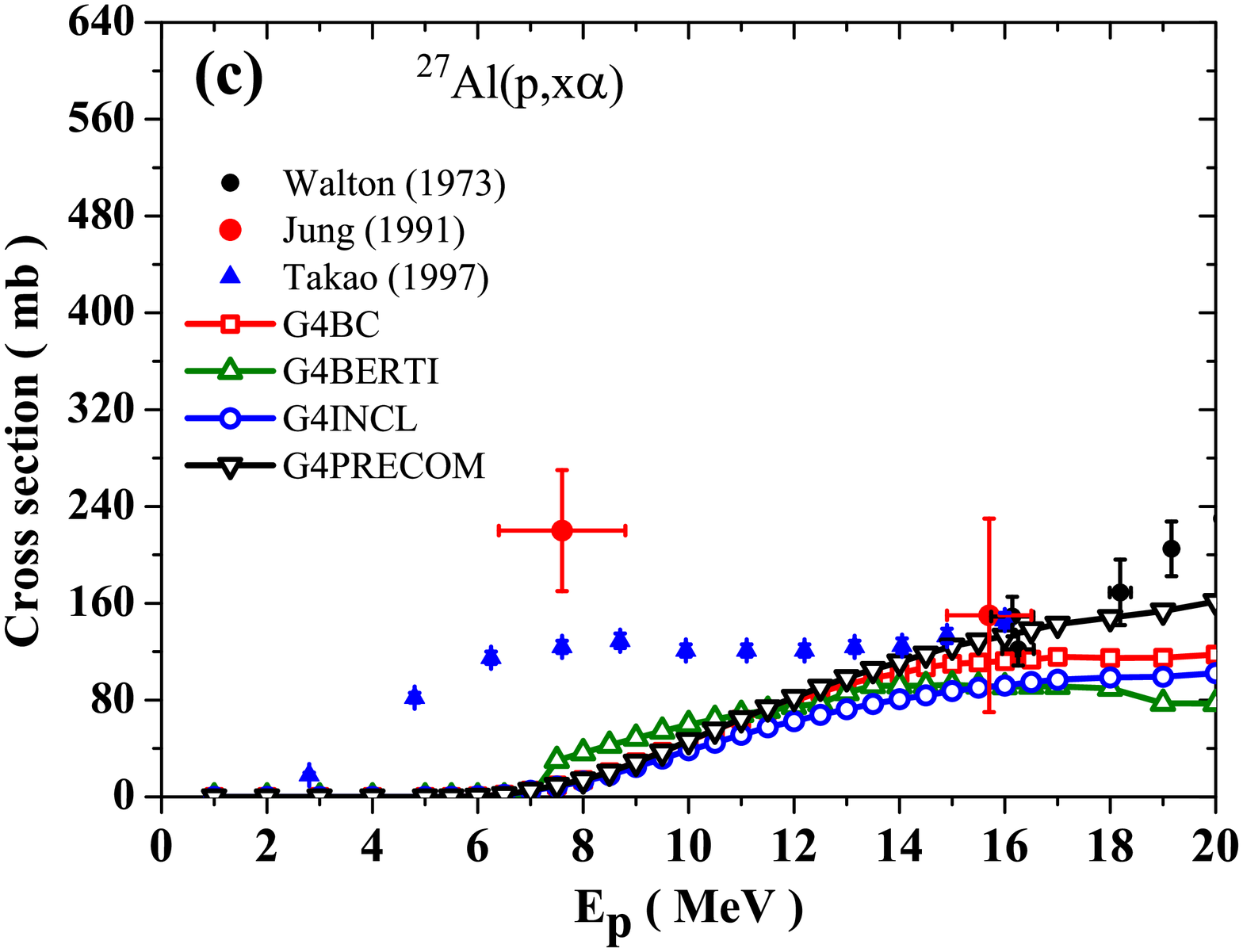, width=2.7in}
\epsfig{file=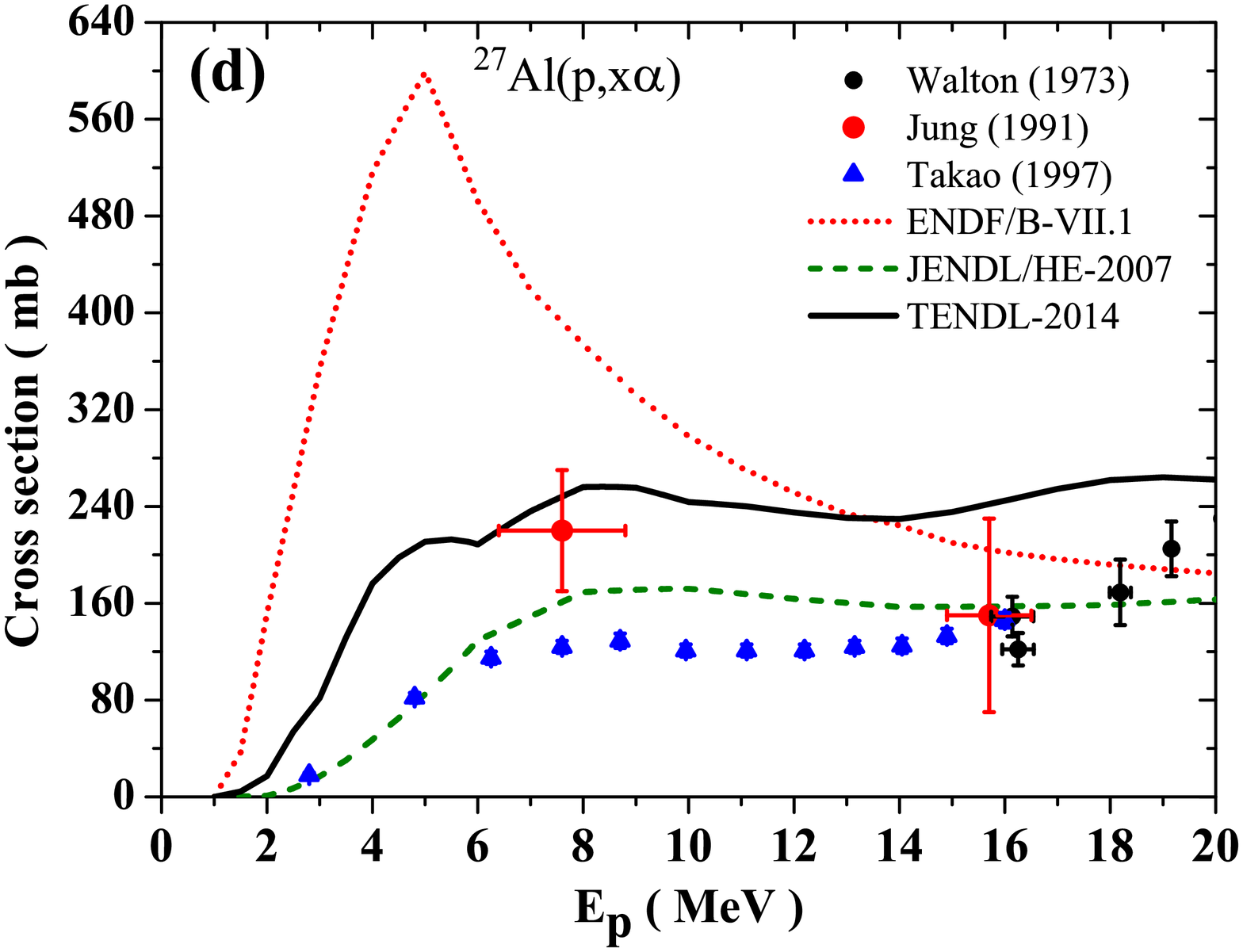, width=2.7in}
\caption{(Color online)
$^{27}$Al(p,xn) reaction cross-sections
are plotted in panel (a) by the hadronic model and panel (b) by the nuclear data model.
(p,x$\alpha$) reaction cross-sections
for the $^{27}$Al are presented in panel (c) and (d)
by the hadronic and nuclear data model, respectively.
See the caption of Fig \ref{fig1}.
for the meaning of symbols and lines.}
\label{fig8}
\end{figure}

For supplementary benchmarking tests,
we also compared the simulated results
by hadronic models of GEANT4
and nuclear data model with experimental cross sections
for $^{27}$Al(p,xn) and $^{27}$Al(p,x$\alpha$) reactions.
As shown in Fig. \ref{fig8} (a) and (c),
all the hadronic models considered underestimate the experimental data,
while the nuclear data model reproduces well the experimental cross section data.

\bibliography{mybibfile}

\end{document}